\documentclass[%
reprint,
floats,amsmath,amssymb,aps
]{article}
\usepackage{jcappub}
\usepackage{float}
\usepackage{graphicx}
\usepackage{dcolumn}
\usepackage{bm}
\usepackage{hyperref}
\usepackage{aas_macros}
\usepackage{tabularx}

\begin{document}

\title{Inside MOND: Testing Gravity with Stellar Accelerations}

\author{Maxwell Finan-Jenkin and}%

\author{Richard Easther}%

\affiliation{%
 Department of Physics, University of Auckland, Private Bag 92019, Auckland, New Zealand\\
}%
\emailAdd{maxwellfj@gmail.com}
\emailAdd{r.easther@auckland.ac.nz}

\date{\today}

\abstract{
We quantify the differences between stellar accelerations in disk galaxies formed in a MONDian universe relative to galaxies with the identical baryonic matter distributions and a fitted cold dark matter halo. In a Milky Way-like galaxy the maximal transverse acceleration is ${\cal {O}}(10^{-9})$ arcseconds per year per decade, well beyond even the most optimistic extrapolations of current capabilities.  Conversely, the maximum difference in the line-of-sight   acceleration is ${\cal {O}}(1)$ centimetre per second per decade at  solar distances from the galactic centre. This level of precision is within reach of plausible future instruments. 
}

\maketitle


\section{Introduction}

Accounting for the dynamical behaviour of the universe at galactic scales and above is one of most enduring problems in fundamental science \cite{thomson,Poincare:1906aaa,Zwicky:1933gu,Zwicky:1937zza}.  Cold dark matter (CDM) is widely believed to be the solution to this conundrum \cite{Peebles:1982ff,Blumenthal:1984bp} and it additionally predicts the observed spectrum of the microwave background and the formation of structure in the expanding universe \cite{1982ApJ...263L...1P}.  The generic CDM candidate is a weakly interacting massive particle  with a mass in the GeV-TeV range \cite{Feng:2010gw}. However, given the huge range of possible CDM scenarios --  from ultralight bosons with masses $10^{25}$ less than that of an electron \cite{Sin:1992bg,Hu:2000ke,Hui:2016ltb} to 100 solar mass primordial black holes \cite{Bird:2016dcv,Carr:2016drx}  -- any given experimental or observational probe explores only a small subset of the total ``theory space'' and it is easy to imagine  CDM variants that are entirely resistant to direct experimental tests. 

Despite its successes, confidence in the CDM hypothesis is thus limited by our inability to answer questions regarding the  detailed properties and underlying physical nature of this novel material. In this context, modified Newtonian dynamics (MOND) \cite{Milgrom:1983ca} remains an active topic. Motivated by the observed velocity profiles in disk galaxies, MOND posits that Newton's laws are altered at very low accelerations. The original {\em ad hoc} proposal has been expressed as a relativistic theory \cite{Bekenstein:2004ca,Skordis:2020eui} and it is argued that MONDian dynamics can account for a variety of otherwise unexplained  small-scale features of galaxies \cite{Famaey:2011kh,McGaugh:2020ppt}. 

This situation is  analogous to  the anomalies in the motion of Mercury and Uranus that perplexed 19th Century astronomers. As is well known, one problem was solved by the discovery of additional matter in the solar system while the other required changes to our understanding of gravitational interactions.  However, the solar system challenges persisted for decades and involved small departures from current models\footnote{Attention was first drawn to the anomalous perihelion precession of Mercury by Le Verrier in the 1850s.} whereas the mismatch between observation and expectation  for galactic dynamics has endured for close to 100 years  and is of order unity.  

As is well known, tests of dark matter proposals  fall into the broad categories of direct detection, accelerator production, and anomalous astrophysical signals generated by dark matter interactions. However, it is increasingly possible that a variety of mechanisms will allow direct measurements of accelerations induced by the galactic potential~ \cite{Silverwood:2018qra,Ravi:2018vqd,Erskine:2019ojg}.  These are  ${\cal{O}}(10^{-10})$ m/s, or roughly ${\cal{O}}(10)$ centimetres per second per decade\footnote{For this reason units of ``cm/s/decade'' provide a useful figure of merit for these discussions.} which corresponds to variations at the parts per billion level in stellar spectra. Detecting such changes requires instruments with both long-term stability and very high dispersion.  The next generation of astronomical instrumentation -- spectrographs coupled to 30 metre class telescopes -- will attain this level of precision, although large systematic effects would need to be isolated~\cite{Silverwood:2018qra}. Direct tests of ``${\bf F}=m{\bf a}$'' would open up new approaches to testing kinematics on kiloparsec scales. In particular, this would allow  probes of MONDian scenarios at a granular level rather than via their broad predictions for galactic morphology and dynamics.  


\begin{figure}[t]
\centerline{\includegraphics[width=15cm]{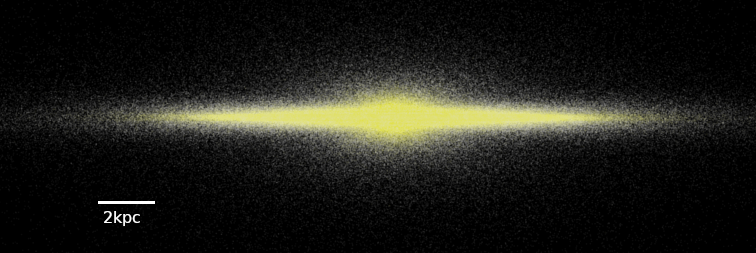}} 
\vspace{.5cm}
\centerline{\includegraphics[width=.7\columnwidth]{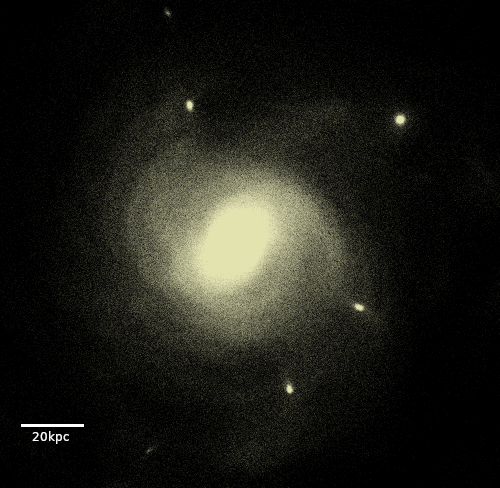}}
\caption{Visualisations of the stellar content of the two largest galaxies drawn from MONDian simulations \cite{Wittenburg:2020abr,Nagesh:2022jof} used in this work. [Top] M4 simulation, viewed side-on; [Bottom]  Top-down view of the 1e11 galaxy, including substructure that evolved during the simulation. Details of  the simulated galaxies are summarized in Table~\ref{tab:galaxies}. }
\label{dense}
\end{figure}

MOND and dark matter predictions for the velocities and thus the accelerations of stars in the galactic disk are in good agreement. This is no coincidence since the MOND hypothesis was  motivated by the observed radial velocity profiles of galactic disks and these are also well-accounted for by the presence of a dark matter halo.  However, a halo -- and its gravitational potential -- is approximately spherically symmetric. Conversely, the MONDian gravitational field of a disk galaxy is sourced  by the baryonic matter and need not be spherically symmetric.  Consequently, there is no reason to expect the vertical acceleration of stars (relative to the plane of the galaxy) to match between MONDian and dark matter scenarios \cite{Silverwood:2018qra,Ravi:2018vqd}.

Our goal here is to  compare the accelerations seen in explicit {MONDian} realisations of disk galaxies to those expected from from the Newtonian potential of a galaxy with the same baryonic component and a spherical dark matter halo chosen to match the accelerations seen in the disk.  We work with the endpoints of MONDian simulations of galaxies formed via spherical collapse \cite{Wittenburg:2020abr} and galaxies initialised in non-expanding backgrounds that pay closer attention to the impact of star-formation and feedback  \cite{Nagesh:2022jof}. These galaxies have masses running from ${\cal{O}}(10^9)$ to ${\cal{O}}(10^{11})$ solar masses, which is close to the baryonic mass of the Milky Way at the upper end. 

The baryonic components of the MONDian and Newtonian galaxies we consider are identical but in general, MONDian galaxies are expected to be morphologically distinct from their Newtonian analogues so our calculation is  an underestimate of the likely differences in the accelerations. Conversely, an asymmetric or triaxial halo could improve the fit. However, we can make a qualitative estimate of the likely differences between the acceleration fields in the two scenarios. Our key finding is that there are systematic differences in the expected radial velocities above and below the disk on the order of $0.5-1$ cm/s/decade which diminish with distance from the galactic centre but increase with the mass of the galaxy. 

This work complements several other approaches to the direct measurement of accelerations in galactic potentials. These include the use of globular clusters as tracers of the overall potential \cite{Quercellini:2008it}, precision pulsar timing \cite{Phillips:2020xmf,Chakrabarti:2020abx,Wang:2021fas,moran2023pulsarbased}, angular accelerations from astrometry \cite{Buschmann:2021izy} and the timing of  eclipsing binaries \cite{Chakrabarti:2021ypi}. Likeiwse, precision measurements of stellar velocities have recently been used to infer accelerations from phase-space methods and contrast MONDian and Newtonian predictions for accelerations out of the disk \cite{2023MNRAS.519.4479Z} and present-day instruments can be used to seek the overall accelerations of stars in the galactic potential \cite{Chakrabarti:2020kco}. Likewise, the trajectories of wide binary stars such as Proxima Centauri \cite{2019arXiv190608264B} or probes to near-interstellar space \cite{2019arXiv190700006B} could test MONDian dynamics. A review of tests of MONDian dynamics is given in Ref.~\cite{Banik:2021woo} and future direct acceleration tests are summarised by Erskine {\em et al.} \cite{Erskine:2019ojg}. Consequently,  considerable effort is being devoted to developing the techniques required to directly test MONDian dynamics.

This structure of this paper is as follows.  In Section~\ref{sec:data} we describe the simulated galaxies.  Section~\ref{sec:fit}  covers the strategies used to fit the dark matter halos to the MONDian results. We describe the differences in the expected line of sight and transverse accelerations in Section~\ref{sec:accel} and conclude in Section~\ref{sec:discuss}.

\section{Simulations and Halo Fits} 
\label{sec:data}

In quasilinear MONDian gravity \cite{Milgrom:2009ee} the Newtonian Poisson solver has a modified source term, 
\begin{equation}
    \nabla^2 \Phi = - \nabla \cdot {\mathbf g} = - \nabla \cdot ( \nu {\mathbf g}_N)
\end{equation}
where ${\mathbf g}$ is the actual gravitational force and  ${\mathbf g}_N$ is the Newtonian expectation for the gravitational force based on the  matter density.  The function $\nu$  interpolates between the Newtonian and Milgromian regimes, and is usually taken to be
\begin{equation}
    \nu = \frac{1}{2} + \sqrt{\frac{1}{4} + \frac{a_0}{|{\mathbf g}_N|}  }
\end{equation}
where $a_0$ is a new fundamental constant. When $|{\mathbf g}_N|$ is large $\nu \rightarrow 1$ and we recover the usual Newtonian limit. 

These equations can be rearranged to yield the distribution of ``phantom'' matter that would be required for the total Newtonian potential to reproduce the MONDian gravitational field, 
\begin{equation}
    \rho_p = (\nu-1) \rho - \frac{1}{4 \pi G a_0} (\nu' \nabla |{\mathbf g}_N |) \cdot  {\mathbf g}_N \, , 
\end{equation}
where $\nu'$ is the derivative of $\nu$ with respect to its argument. The form of $\rho_p$ is fully specified by $\rho$ so its overall form is interesting but not necessarily informative. Conversely, dark matter is a genuinely independent contribution to the overall density and thus the gravitational potential. In particular, in a spiral galaxy the symmetries of the dark and baryonic components are very different so there is no reason to expect that $\rho_p$ will exactly reproduce $\rho_{{\mathrm \footnotesize DM}}$ in a spiral galaxy.

\begin{table}
\begin{center}
\begin{tabularx}{.85 \textwidth}{ccccccc}
\hline\hline
   Name  & $N$ & $M_\star$ ($10^9$M$_\odot$) & $M_g$ ($10^9$M$_\odot$) & $R_\mathrm{eff}$ (kpc) & $L$ (kpc) &Source \\ \hline
    M1 & $1.2\times10^5$ & 5.215 & 1.190 & 6.608 & 200 & \cite{Wittenburg:2020abr} \\
    1e10 & $2.0\times10^6$ & 11.01 & 2.800 & 22.24 & 400 & \cite{Nagesh:2022jof}\\
    M2 & $3.2\times10^5$ & 21.61 & 1.130 & 5.005 & 400  & \cite{Wittenburg:2020abr}  \\
    M3 & $4.2\times10^5$ & 21.61 & 2.378 & 8.782& 400  & \cite{Wittenburg:2020abr}  \\
    M4 & $1.2\times10^6$ & 95.62 & 4.420 & 17.09 & 400 &  \cite{Wittenburg:2020abr} \\
    1e11 & $2.2\times10^6$ & 122.4 & 10.6 & 47.53 & 400 &\cite{Nagesh:2022jof}\\
    \hline\hline
\end{tabularx}
\end{center}
\caption{\label{tab:galaxies} Summary of simulated disk galaxies; each galaxy has $N$ stars, stellar mass $M_\star$, gas-mass $M_g$ and effective radius $R_\mathrm{eff}$,  containing 95\% of the stellar mass. The overall simulation box is a volume of $L^3$.}
\end{table}

\begin{figure}[tb]
\centerline{\includegraphics[width= .9 \columnwidth]{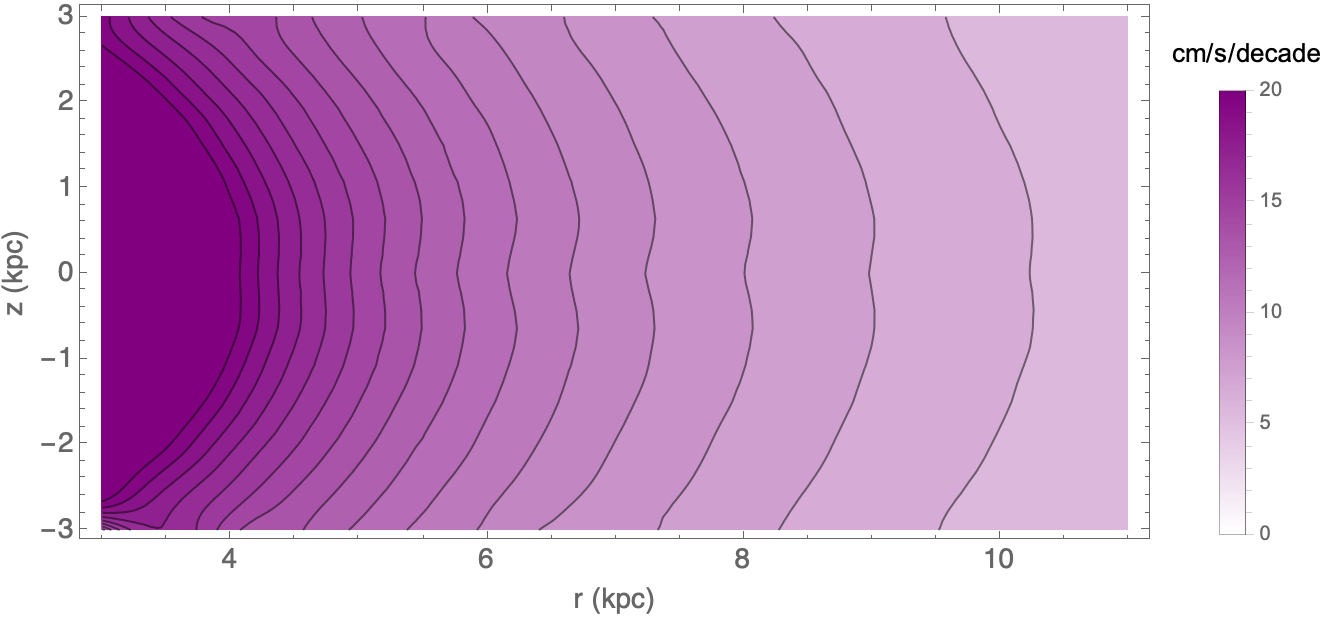}}  
\caption{  Radially averaged norm of MONDian acceleration for the M4 galaxy.  
} 
\label{accelsmond}
\end{figure}

It is impossible to create fully analogous pairs of galaxies from initial conditions in the two frameworks given that we want to match their endpoints and they have will different evolutionary histories. Consequently, we take realisations of galaxies from MONDian simulations and compute the radial density distribution of the spherical dark matter halo needed to reproduce the observed accelerations in the disk. We then compare the  accelerations in the two scenarios, both parallel and transverse to the line of sight.  This is not the only possible approach but it is an ``apples to apples'' comparison.

The simulated galaxies are listed in Table~\ref{tab:galaxies}, drawn from Refs~\cite{Wittenburg:2020abr} and \cite{Nagesh:2022jof}. Both papers use a modified version  \cite{Lughausen:2014ala} of the RAMSES code \cite{Teyssier:2001cp}. In the former, simulations begin as an isolated, collapsing overdensity in an expanding universe that initially contains only gas but undergoes star formation and,   after gravitational collapse,  models the effects of supernovae and feedback. The latter simulations give greater attention to star formation and detailed feedback dynamics but are initialised with the galaxies in place.  Representative simulated galaxies are depicted in Fig.~\ref{dense}. The simulations of Ref~\cite{Wittenburg:2020abr} show that gravitational collapse in a MONDian scenario can  yield exponential disk galaxies with radially symmetric stellar distributions. Including more detailed internal dynamics leads to the formation of substructure \cite{Nagesh:2022jof} in the outer parts of the galaxy.  The larger halos have $N \sim 10^6$  mass points. 
 
In all scenarios the final gas fraction is a subdominant but still significant component ($\lesssim 20$\%) of the total mass and is typically collapsed to a thin disk. For the M4 case gas typically makes a contribution on the order of  0.01 cm/s/decade, less than 1\% of the overall gravitational acceleration of roughly 10~cm/s/decade experienced by stars near the disk  but we include the gravitational fields of both components in the analysis that follows. The radially averaged acceleration of the disk of the M4 galaxy is shown for reference in Fig.~\ref{accelsmond}. We use this simulation  as our canonical example in what follows and focus on the region of the galaxy away from the core.

\section{Halo Fitting}
\label{sec:fit}

We build a synthetic dark matter halo that reproduces the stellar accelerations seen in the disk of the simulated galaxies. Given this ``synthetic'' halo we then compare the accelerations of stars above and below the disk (which are purely gravitational, unlike the gas dynamics)  to the MONDian accelerations extracted from the simulations.

\begin{figure}[tb]
\centerline{\includegraphics[width= .7 \columnwidth]{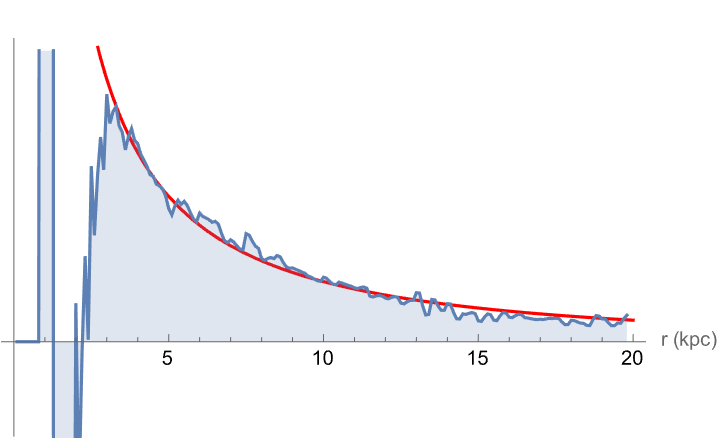}}
\caption{ Inferred dark matter density profile  for the M4 galaxy and fitted NFW profile (using the inferred densities at  $r \ge 3$~kpc), as  function of radial distance $r$. 
} 
\label{onion ring}
\end{figure}

We build the synthetic halo as an ``onion'', or a series of concentric spherical shells. The specific procedure is as follows: 
\begin{enumerate}
\item Select all the ``stars'' in a rectangular toroidal region of height $h$ and radius $r$ to $r+\delta r$, relative to the galactic centre of mass.  
\item Calculate the Newtonian gravitational acceleration on each star from the overall mass distribution.
\item Calculate the average difference between the Newtonian accelerations   and the MONDian accelerations for each rectangular toroidal bin.
\item Compute the mass of the spherical dark matter shell that accounts for this difference. 
\end{enumerate}  
This strategy guarantees that the disk accelerations in the MONDian simulation match those of our synthetic dark matter galaxy. We use $d=10$~pc and $\delta r=250$~pc but our results are insensitive to these specific choices.

The density profile inferred for the M4 galaxy is shown in Fig. \ref{onion ring}, along with a fit to the NFW profile, 
\begin{equation} 
\rho(r) = \frac{\rho_0}{\frac{r}{r_s}\left(1+\frac{r}{r_s}\right)^2} \, .
\end{equation}
For the M4  case the scale radius is roughly $45$~kpc. The fit is very poor for radial distances below $3$~kpc -- the inferred density is actually negative at some points, which reflects the fact that MONDian galaxies have quite different central dynamics from their CDM counterparts.  In what follows we use a smoothed onion model to compute the synthetic halo mass as a function of radius rather than the  NFW fit and we restrict attention to regions of the galaxy in which the NFW fit is reliable.

We calculate the Newtonian acceleration of each star by direct summation, which has an $N^2$ computational cost. Solving Poisson's equation and then inferring the accelerations from the potential would  be more time-efficient from a computational perspective but since it only needed to be performed a relatively small number of times the brute-force calculation is tractable and convenient.

\begin{figure}[tb]
\centerline{\includegraphics[width= .9 \columnwidth]{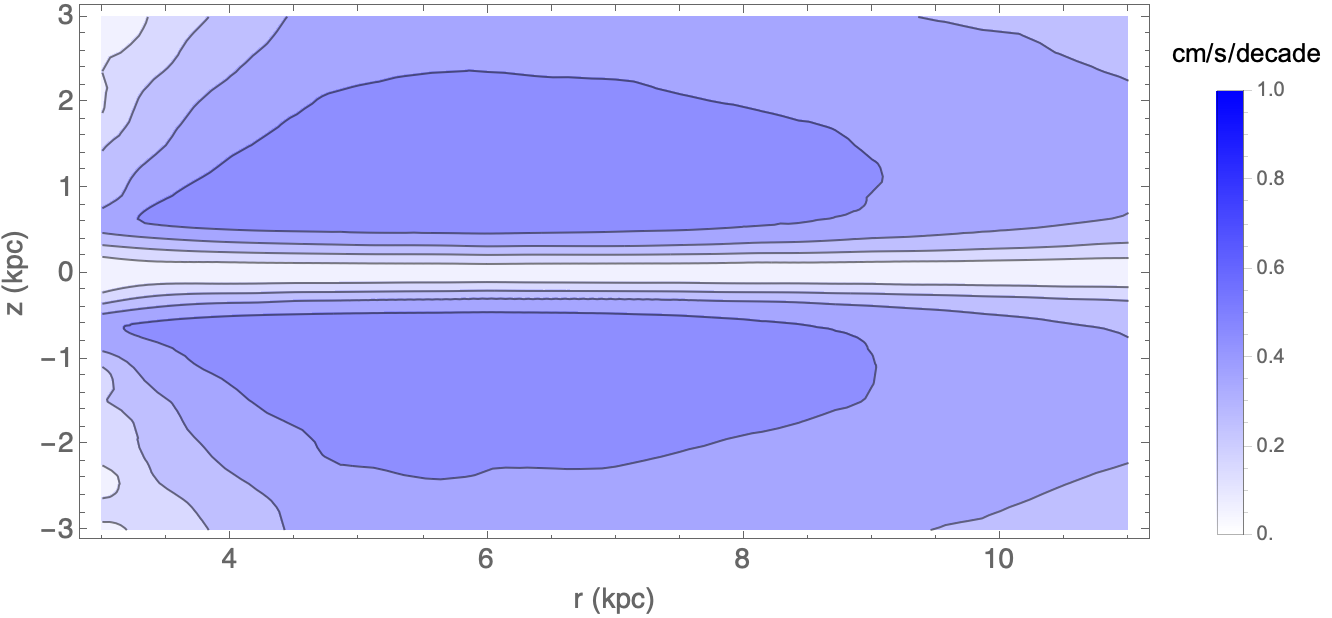}} 
\caption{   Norm of the radially averaged difference between the   MONDian acceleration and acceleration predicted by the synthetic halo. Note that the difference vanishes in the plane, by construction.
} 
\label{accels}
\end{figure}

We considered both Plummer potentials and excluding adjacent stars below a fixed cutoff length to guard against coincidentally adjacent $N$-body particles experiencing  unphysical accelerations due to the  spurious granularity of  the simulated galaxies. In practice we used a hard exclusion radius of 130pc but the dependence on this value or the specific smoothing choice is very weak. Fig.~\ref{accels} shows the radially averaged differences between the MONDian acceleration and those with a synthetic halo for the canonical M4 galaxy. As expected the difference vanishes when the vertical displacement $z\approx 0$ which serves as a check on our halo fit. The maximum difference is on the order of 0.5~cm/s/decade.

With the galaxy at the origin of the Cartesian coordinate system the distance between an observer in the plane at the same angular position and a star at  $(x,y,z)$ is
\begin{equation}
{\mathbf r}' =  \left( x \left( 1- \frac{r}{\sqrt{x^2 + y^2}}\right) ,\left( 1- y  \frac{r}{\sqrt{x^2 + y^2}}\right) ,z \right) \, . 
\end{equation}
The star has an acceleration ${\mathbf a}$ which splits into line of sight (spatial) and transverse (angular) components $a_{\text{LoS}}$ and $a_{\perp}$ respectively, via 
\begin{equation}
    a_{\text{LoS}} = \frac{ {\mathbf r}' \cdot {\mathbf {\mathbf a}}}{|{\mathbf r}'|}
\end{equation}
and 
\begin{equation}
    a_{\perp} = \frac{\sqrt{ |{\mathbf a}|^2 -   a_{\text{LoS}}^2}}{|{\mathbf r}'|} \,  .
\end{equation}
We assume radial symmetry and  bin the stellar accelerations in the $\{r,z\}$ plane to obtain our results, where $r=\sqrt{x^2+y^2}$ Fig.~\ref{diff7} shows the projected acceleration seen by an inertial observer located 7~kpc from the centre of the M4 galaxy as a representative example. 

Note that these accelerations are defined relative to an inertial reference frame. In practice, any actual measurement would be made the accelerated frame that moves with the Sun and we can similarly define the differential accelerations. 

\section{Acceleration Differences: MOND v CDM}
\label{sec:accel}

\begin{figure}[tb]
\centerline{\includegraphics[width=  \columnwidth]{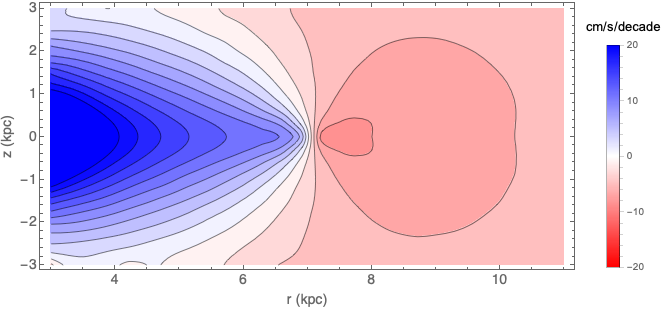}} 
\mbox{}\\
\centerline{\includegraphics[width= \columnwidth]{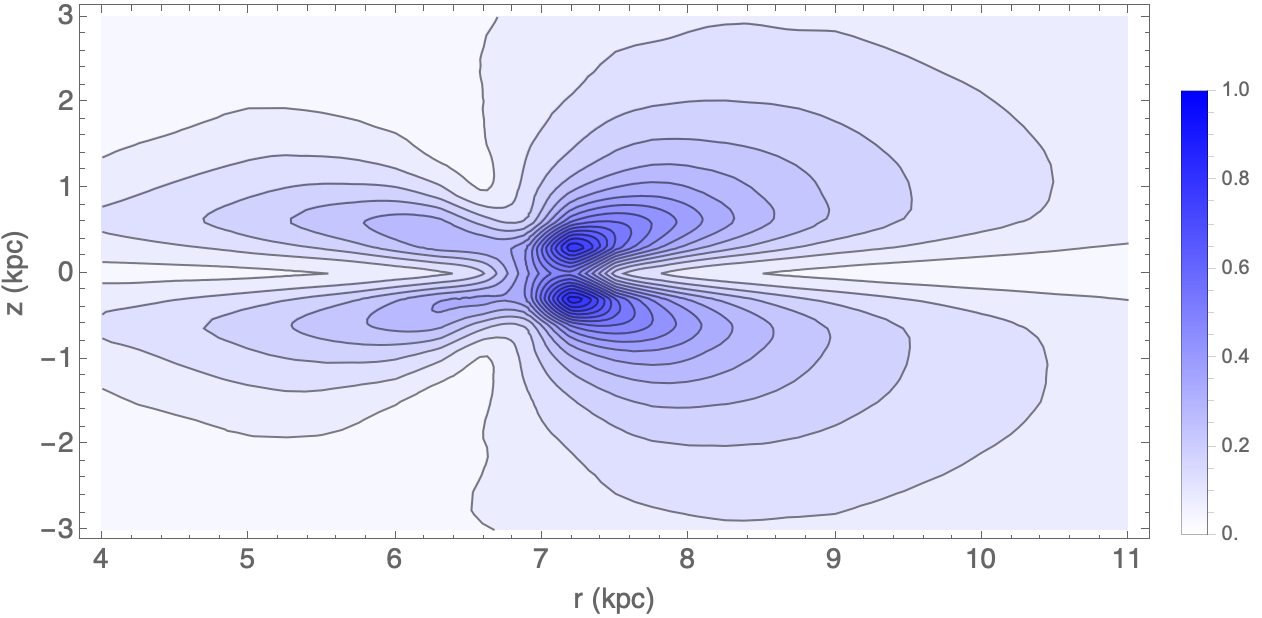}}
\caption{Difference in line of sight [top] and transverse accelerations [bottom] for an observer 7 kpc from the center of the M4 galaxy; $a_\perp$ has units of nano-arcseconds/year/decade.}
\label{diff7}
\end{figure}

Define ${\mathbf \Delta}={\mathbf a}_{\text{N}}-{\mathbf a}_{\text{M}}$, the (vector) difference between Newtonian and MONDian accelerations. We verify that our fitted halo behaves as expected by plotting the radially averaged value of $|{\mathbf \Delta}|$ in Fig.~\ref{accels}. The difference vanishes in the plane and the typical maximum difference in the accelerations is on the order of 0.5 cm/s/decade. 

In principle, both transverse and radial accelerations could be observed. Fig.~\ref{diff7} shows  $\Delta_{\mathrm{LoS}}$ and $\Delta_{\perp}$ for an inertial observer 7~kpc from the centre of the $M4$ galaxy. However, as expected from the analysis of Ref.~\cite{Silverwood:2018qra}, $\Delta_\perp$ is very small,  typically less than 1 nanoarcsecond/year/decade and this is common across all the simulated galaxies. It is unlikely that this level of acceleration would be observable, even with optimistic extrapolations of future astrometric surveys.

\begin{figure}[!ht]
\begin{center}
\vspace{1.2cm}
\includegraphics[width=.99\textwidth]{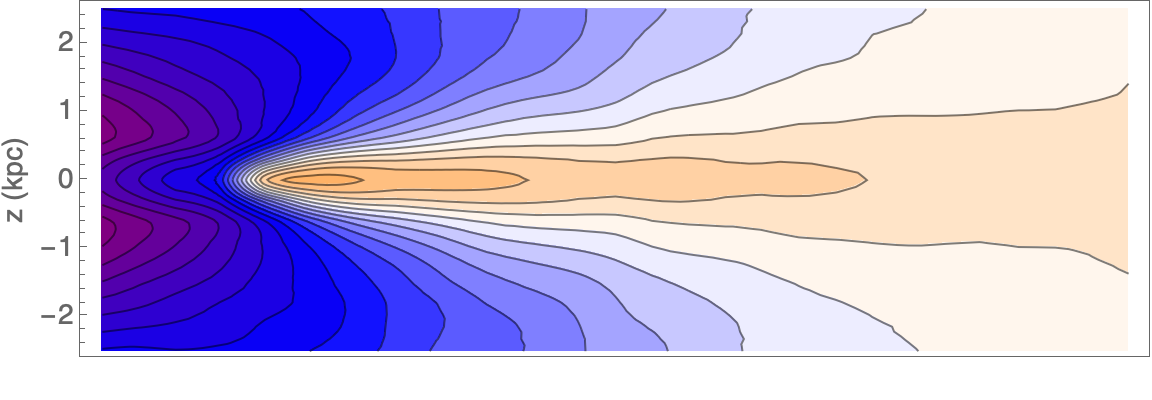} \\
\includegraphics[width=.99\textwidth]{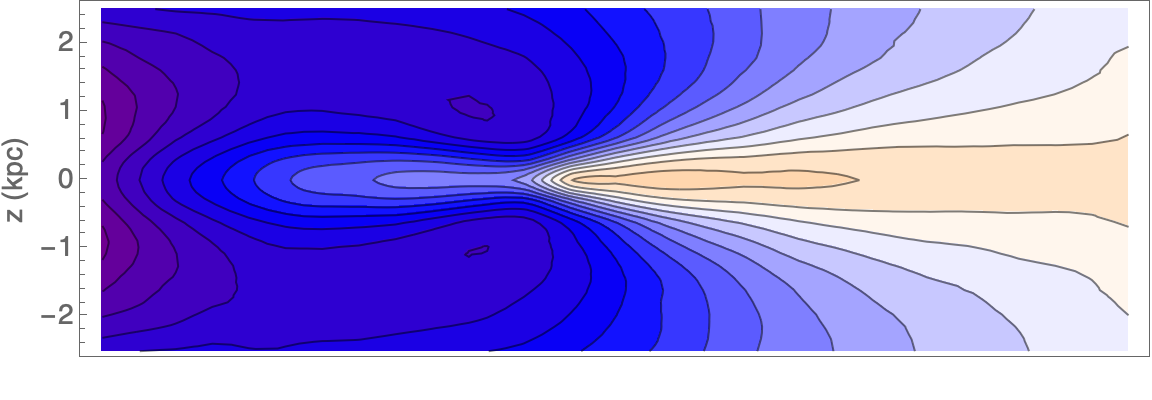} \\
\includegraphics[width=.99
\textwidth]{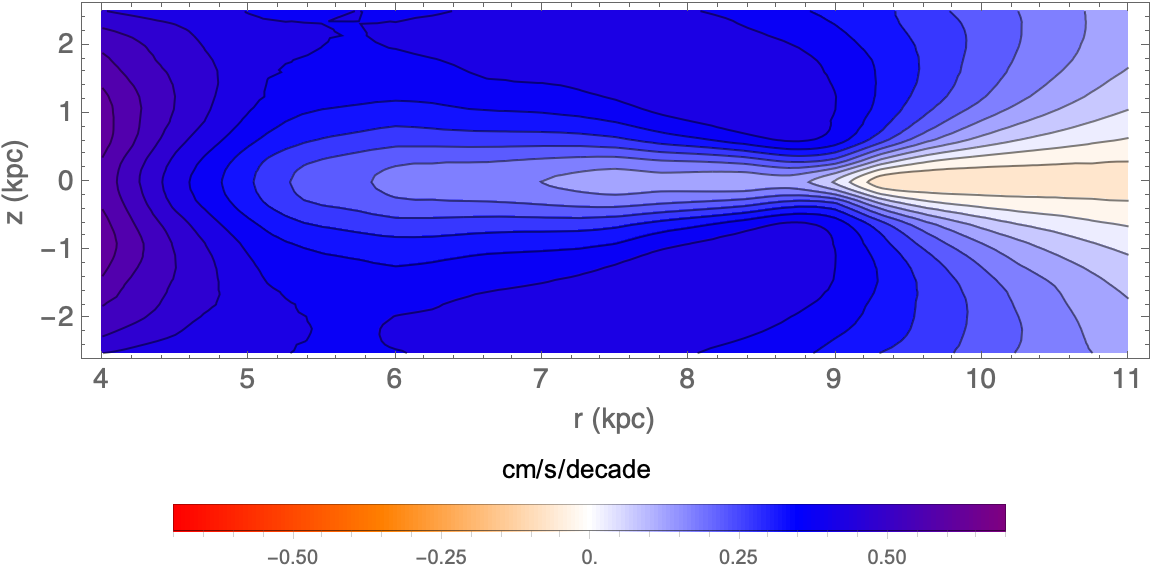}
\end{center}
\caption{Difference in line-of-sight acceleration between MOND and dark matter  for observers 5, 7 and 9 kpc (top to bottom) from the center of the M4 galaxy.}\label{losdiff}
\end{figure}

Fig. \ref{losdiff} shows the line-of-sight differences in the expected acceleration for observers at 5, 7 and 9 kpc from the centre of the M4 galaxy.   The largest differences are found above and below the plane and are of the order 0.5 cm/s/decade. Larger values are seen toward the galactic centre which reflects the breakdown of the match between the NFW profile and MONDian dynamics in this region. Measurements of this precision could be  possible with reasonable extrapolations of instruments now under construction, although it would be an heroic achievement to actually extract individual stellar accelerations of this size given the  systematics involved. Conversely, a campaign involving multiple stars would facilitate a statistical detection with lower individual precision \cite{Ravi:2018vqd}.

\begin{figure}[t]
\includegraphics[width=\columnwidth]{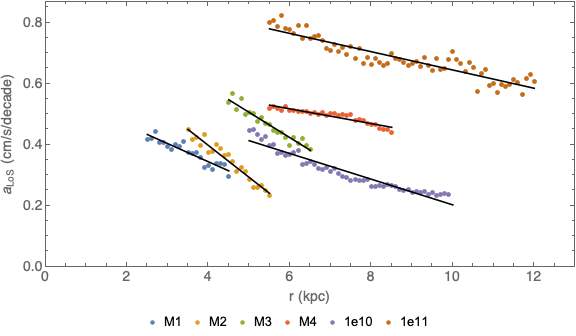}
\caption{The value $\Delta_{\text{LoS}}$ in a region 0.9 kpc vertically out of the plane is plotted as a function of radius for each of the simulated galaxies, for observers at $z=0$. }\label{variation with radius}
\end{figure}

Fig. \ref{variation with radius} shows $\Delta_{\text{LoS}}$ as a function of radius at $z= \pm 0.9$  (the rough location of the maximal value) at the same angular displacement as the observer  for all six simulated galaxies. We see two clear trends -- the maximal value  of $\Delta_{\text{LoS}}$ drops roughly linearly with the distance from the centre in a given galaxy, and the total effect grows with the overall galactic mass, albeit slowly. The total mass range is of the order a factor of 20 and the largest galaxy (1e11) is roughly a factor of two smaller than the bayonic mass of the Milky Way.

\begin{figure}[t]
\includegraphics[width=\columnwidth]{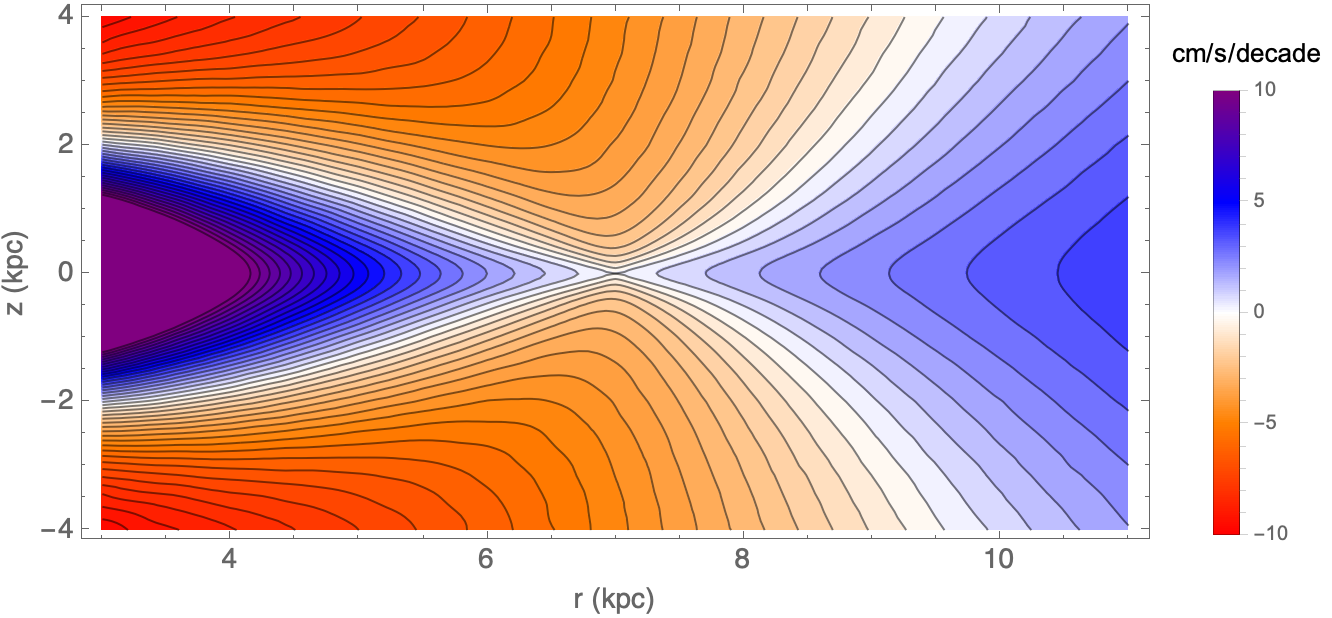} \\
\mbox{}\\
\includegraphics[width=\columnwidth]{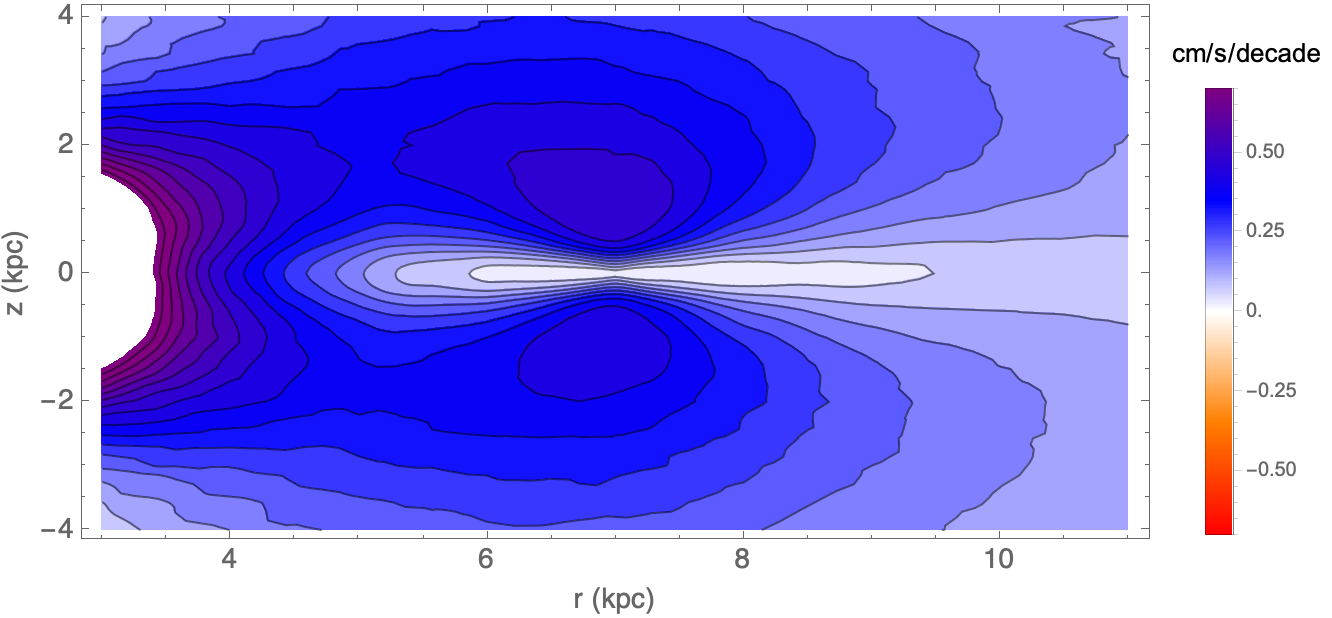}
\caption{[Top] $a_{\text{LoS}}$ seen by an accelerated observer 7 kpc from the centre of the M4 galaxy; [Bottom]  $\Delta_{\text{LoS}}$ for by an accelerated observer at the same position; note the different scales on the two plots} \label{excl1}
\end{figure}

As pointed out earlier, we have been plotting the acceleration differences relative to an inertial observer that are due solely to the differences between Newtonian and MONDian gravity, for the same baryonic source distributions of stars and gas. However, any realistic observation measures a relative acceleration. In Fig.~\ref{excl1} and we plot the relative acceleration and the relative acceleration difference seen from within the canonical M4 galaxy. This can be compared with the top panel of Fig.~1 from Ref.~\cite{Silverwood:2018qra} which shows the same plot for the Milky Way potential. The maximal relative acceleration differences are directly above and below the observer, relative to the galactic plane.

\section{Conclusion}
\label{sec:discuss}

We have looked at the differences between the accelerations seen MONDian and Newtonian realisations of the same galaxy. We have confirmed that while these are very small they will be potentially observable in foreseeable instruments.  

This was achieved by taking galaxies formed in MONDian simulations and finding dark matter distributions that would duplicate the  rotation seen in the plane of the disk. We then calculate the difference in the accelerations that would be measured in the MONDian and Newtonian realisations. 
For an observer at a roughly earth-like position in a Milky Way sized galaxy the maximal difference will be a little less than 1 cm/s/decade in the line of sight.  This level of precision is potentially achievable  by the next generation of instruments over timescales of a decade. The largest accelerations are directly above and below the observer's position in the disk. 

That said, a number of caveats apply. Many large systematics will need to be isolate to attribute sub-cm/s changes in velocity to the galactic potential.  These include astrometric accelerations (whereby proper motion changes the division between between transverse and line-of-sight accelerations), stellar ``jitter'' and reflex velocities induced by long-period planetary companions \cite{Silverwood:2018qra}.  Sophisticated observing strategies would be needed to control for these issues, as well as reaching the raw level of precision to make the distinction between different expectations for galactic accelerations. Conversely we have not attempted to account for any systematic morphological differences between the baryonic content of MONDian and Newtonian galaxies, and these effects  might generally be expected to increase the differential accelerations.  

This work is based on simulations that reflect an early level of maturity in the overall analysis of galaxy formation in MONDian gravity.\footnote{This point was made to us by the authors of Ref.~\cite{Wittenburg:2020abr} and \cite{Nagesh:2022jof}. } Consequently, the predicted distributions of baryonic matter in these scenarios may change as more sophisticated codes are developed. In particular the matter component in these simulations is purely baryonic so there is more scope for stellar feedback and subgrid physics to change their overall morphology that is the case with CDM.
Conversely, because we have fit synthetic halos to  baryonic mass distributions derived from MONDian simulations we have removed any variation due to morphological differences between the MOND and Newtonian scenarios.

Direct observation of stellar accelerations provides a fascinating target for next generation observatories. That said, our goal here is not to make definitive predictions for the differences between MONDian and dark matter dominated galaxies. Rather our aim was to clarify the magnitude of the differences  we can expect and  thus set benchmarks for the precision needed for future measurements of stellar accelerations to test MONDian gravity. Achieving this would break new ground for  galactic astrophysics and fundamental tests of gravitational dynamics at kiloparsec scales.

\begin{acknowledgments} 
We are grateful to Srikanth Nagesh, Nils Wittenburg and collaborators in the Stellar Populations and Dynamics Research  
(SPODYR) group for making the results of their simulations available and for useful feedback on our results. We thank Hamish Silverwood for a number of  conversations on this topic.  RJME acknowledges support of the Marsden Fund  managed through Royal Society Te Ap\={a}rangi. 
\end{acknowledgments}


\providecommand{\noopsort}[1]{}\providecommand{\singleletter}[1]{#1}%

\providecommand{\href}[2]{#2}\begingroup\raggedright\endgroup

\end{document}